\documentclass{article}
\usepackage{epsfig}
\usepackage{latexsym}
\begin{document}

\author{
R. Vilela Mendes \\
Grupo de F\'\i sica-Matem\'atica\\
Complexo Interdisciplinar, Universidade de Lisboa \\
Av. Gama Pinto, 2, 1699 Lisboa Codex Portugal\\
e-mail: vilela@alf4.cii.fc.ul.pt}
\title{\bf Dynamics of networks and applications}
\date{}
\maketitle

\begin{abstract}
A survey is made of several aspects of the dynamics of networks, with
special emphasis on unsupervised learning processes, non-Gaussian data
analysis and pattern recognition in networks with complex nodes.
\end{abstract}

\tableofcontents

\section{Recurrent networks. Dynamics and applications}

There are three large classes of neural networks. One is the class of {\it %
multilayered feedforward networks} which, through supervised learning, are
used to approximate nonlinear functions. The second is the class of {\it %
relaxation networks} with symmetric synaptic connections, like the Hopfield
network. Under time evolution the relaxation networks evolve to fixed points
which, in successful applications, are identified with memorized patterns.
In the last class one includes all networks with arbitrary connections which
are neither completely feedforward nor symmetric. They are called {\it %
recurrent networks}.

Recurrent networks exhibit a complex variety of temporal behavior and are
increasingly being proposed for use in many engineering applications\cite
{Sato1} \cite{Sato2} \cite{Pearlmutter} \cite{Lin}. In contrast to
feedforward or relaxation networks, the rich dynamical structure of
recurrent networks makes them a natural choice to process temporal
information. Even for the learning of nonlinear functions the feedback
structure improves the reproduction of discontinuities or large derivative
regions\cite{Jones}. Also, in some cases, the feedback structure is a way to
enhance, through a choice of architecture, the sensitivity of the network to
particular features to be detected\cite{Amorim}. The learning algorithms
used for feedforward networks may be generalized to the recurrent case\cite
{Pineda} \cite{Borges1} \cite{Borges2} \cite{Amorim}. Some of this will be
covered by another speaker at this conference.

Feedback loops and coexistence of quiescent, oscillatory and chaotic
behavior are also present in biological systems and only the recurrent
networks are appropriate models for these phenomena\cite{Freeman1} \cite
{Amit1}.

The whole field of recurrent networks, both for engineering and biological
applications, is developing in many directions. Their global analysis is
rather involved\cite{Hertz}. Their characterization as dynamical systems is
sharpened by a decomposition theorem. Continuous state - continuous time
neural networks, as well as many other systems\cite{Grossberg}, may be
written in the Cohen-Grossberg\cite{Cohen} form 
\begin{equation}
\frac{dx_{i}}{dt}=a_{i}(x_{i})\left\{
b_{i}(x_{i})-\sum_{j=1}^{n}w_{ij}d_{j}(x_{j})\right\}  \label{1.1}
\end{equation}
Dynamical systems of this type have a decomposition property. Define

\begin{equation}
\begin{array}{ccl}
w_{ij} & = & w_{ij}^{(S)}+w_{ij}^{(A)} \\ 
w_{ij}^{(S)} & = & \frac{1}{2}\left( w_{ij}+w_{ji}\right) \\ 
w_{ij}^{(A)} & = & \frac{1}{2}\left( w_{ij}-w_{ji}\right) \\ 
V^{(S)} & = & -\sum_{i=1}^{n}\int^{x^{i}}b_{i}(\xi _{i})d_{i}^{^{\prime
}}(\xi _{i})d\xi ^{i}+\frac{1}{2}%
\sum_{j,k=1}^{n}w_{jk}^{(S)}d_{j}(x^{j})d_{k}(x^{k}) \\ 
H & = & \sum_{i=1}^{n}\int^{x^{i}}\frac{d_{i}(\xi _{i})}{a_{i}(\xi _{i})}%
d\xi _{i}
\end{array}
\label{1.2}
\end{equation}
Then we have the following

{\it Theorem} \cite{Vilela1} If $a_{i}(x_{i})/d_{i}^{^{^{\prime }}}(x_{i})>0$
$\forall x,i$ and the matrix $w_{ij}^{(A)}$ has an inverse then the vector
field $\stackrel{\bullet }{x_{i}}$ in Eq.(\ref{1.1}) decomposes into one
gradient and one Hamiltonian components, $\stackrel{\bullet }{x_{i}}=%
\stackrel{\bullet }{x_{i}}^{(G)}+\stackrel{\bullet }{x_{i}}^{(H)}$, where 
\begin{equation}
\begin{array}{ccl}
\stackrel{\bullet }{x_{i}}^{(G)} & = & -\frac{a_{i}(x_{i})}{d_{i}^{^{\prime
}}(x_{i})}\frac{\partial V^{(S)}}{\partial x_{i}}=-\sum_{j}g^{ij}(x)\frac{%
\partial V^{(S)}}{\partial x_{j}} \\ 
\stackrel{\bullet }{x_{i}}^{(H)} & = & 
-\sum_{j}a_{i}(x_{i})w_{ij}^{(A)}(x)a_{j}(x_{j})\frac{\partial H}{\partial
x_{j}}=\sum_{j}I^{ij}(x)\frac{\partial H}{\partial x_{j}}
\end{array}
\label{1.3}
\end{equation}
and 
\begin{equation}
\begin{array}{ccl}
g^{ij}(x) & = & \frac{a_{i}(x_{i})}{d_{i}^{^{\prime }}(x_{i})}\delta ^{ij}
\\ 
\omega _{ij}(x) & = & -a_{i}(x_{i})^{-1}\left( w^{(A)-1}\right)
_{ij}(x)a_{j}(x_{j})^{-1}
\end{array}
\label{1.4}
\end{equation}
$\left( \omega _{ij}I^{jk}=\delta _{i}^{k}\right) $. $g^{ij}(x)$ and $\omega
_{ij}(x)$ are the components of the Riemannian metric and the symplectic
form.

{\it Proof} : The decomposition follows by direct calculation from (\ref{1.1}%
) and (\ref{1.2}). The conditions on $a_{i}(x_{i})$, $d_{i}^{^{\prime
}}(x_{i})$ and $w^{(A)}$ insure that $g$ is a well defined metric and $%
\omega $ is non-degenerate. Indeed let $v$ be a vector such that $%
\sum_{i}v^{i}\omega _{ij}=0$. Then 
\[
0=\sum_{ij}v^{i}\omega _{ij}a_{j}(x_{j})w_{jk}^{(A)}(x)=-\frac{v^{k}}{%
a_{k}(x_{k})} 
\]
would imply $v^{k}=0$ $\forall k$. That $\omega $ is a closed form follows
from the fact that $\omega _{ij}$ depends only on $x_{i}$ and $x_{j}$. $%
\Box $

The identification, in the system (\ref{1.1}), of just one gradient and one
Hamiltonian component with explicitly known potential and Hamiltonian
functions, is a considerable simplification as compared to a generic
dynamical system. Recall that in the general case, although such a
decomposition is possible locally\cite{Vilela2}, explicit functions are not
easy to obtain unless one allows for one gradient and $n-1$ Hamiltonian
components. Notice that the decomposition of the vector field does not
decouple the dynamical evolution of the components. In fact it is the
interplay of the dissipative (gradient) and the Hamiltonian component that
leads, for example, to limit cycle behavior.

For the case of symmetric connections $w_{ij}=w_{ji}$ one recovers the
Cohen-Grossberg result\cite{Cohen} that states that a symmetric system of
the type of Eq.(\ref{1.1}) has a Lyapunov function $V^{(S)}$ of which
Hopfield's\cite{Hopfield} ''energy'' function is a particular case. For the
symmetric case the existence of a Lyapunov function guarantees global
asymptotic stability of the dynamics. However not all vector fields with a
Lyapunov function are differentially equivalent to a gradient field.
Therefore the fact that a gradient vector is actually obtained gives
additional information, namely about structural stability of the model.

A necessary condition for structural stability of the gradient vector field
is the non-degeneracy of the critical points of $V^{(S)}$, namely $\det
\left\| \frac{\partial ^{2}V^{(S)}}{\partial x_{i}\partial x_{j}}\right\|
\neq 0$ at the points where $\frac{\partial V^{(S)}}{\partial x_{i}}=0$. In
a gradient flow all orbits approach the critical points as $t\rightarrow
\infty $. If the critical points are non-degenerate then the gradient flow
satisfies the conditions defining a Morse-Smale field, except perhaps the
transversality conditions for stable and unstable manifolds of the critical
points. However because Morse-Smale fields are open and dense in the set of
gradient vector fields, any gradient flow with non-degenerate critical
points may always be C$^{1}$-approximated by a (structurally stable)
Morse-Smale gradient field. Therefore given a symmetric model of the type (%
\ref{1.1}), the identification of its gradient nature provides a easy way to
check its robustness as a physical model.

As an example of the decomposition applied to a biological model consider
the Wilson-Cowan model of a neural oscillator without refractory periods, in
the antisymmetric coupling case considered by most authors 
\begin{equation}
\begin{array}{lll}
\stackrel{\bullet }{x_{1}} & = & -x_{1}+S\left( \rho
_{1}+w_{11}x_{1}+w_{12}x_{2}\right) \\ 
\stackrel{\bullet }{x_{2}} & = & -x_{2}+S\left( \rho
_{2}+w_{21}x_{1}+w_{22}x_{2}\right)
\end{array}
\label{1.5}
\end{equation}
with $w_{12}=-w_{21}$ and $S$ is the sigmoid function $\left(
1+e^{-x}\right) ^{-1}$ . Changing variables to 
\[
z_{i}=\rho _{i}+\sum_{i=1}^{2}w_{ij}x_{j} 
\]
one obtains 
\begin{equation}
\begin{array}{lll}
\stackrel{\bullet }{z_{1}} & = & -\frac{\partial V}{\partial z_{1}}+w_{12}%
\frac{\partial H}{\partial z_{2}} \\ 
\stackrel{\bullet }{z_{2}} & = & -\frac{\partial V}{\partial z_{2}}-w_{12}%
\frac{\partial H}{\partial z_{1}}
\end{array}
\label{1.6}
\end{equation}
with 
\[
\begin{array}{lll}
V & = & \frac{1}{2}\sum_{i}\left\{ z_{i}^{2}-\rho _{i}z_{i}+w_{ii}\log
\left( 1-S(z_{i})\right) \right\} \\ 
H & = & \sum_{i}\log \left( 1-S(z_{i})\right)
\end{array}
\]
The model is completely described by these functions, the bifurcation sets%
\cite{Frank}, for example, being characterized by $\triangle V=0$ for
Andronov-Hopf bifurcations and by 
\[
\frac{\partial ^{2}V}{\partial z_{1}^{2}}\frac{\partial ^{2}V}{\partial
z_{2}^{2}}+w_{12}^{2}\frac{\partial ^{2}H}{\partial z_{1}^{2}}\frac{\partial
^{2}H}{\partial z_{1}^{2}}=0 
\]
for saddle-node bifurcations.

\section{Unsupervised learning in generalized networks and the processing of
non-Gaussian signals}

\subsection{Unsupervised leaning in general networks}

Doyne Farmer\cite{Farmer} has shown that there is a common mathematical
framework where neural networks, classifier systems, immune networks and
autocatalytic reaction networks may be treated in a unified way. The general
model in which all these models may be mapped looks like a neural network
where, in addition to the node state variables ($x_{i}$) and the connection
strengths ($W_{ij}$), there is also a node parameter ($\theta _{i}$) with
learning capabilities (Fig.1).
\begin{figure}[htb]
\begin{center}
\psfig{figure=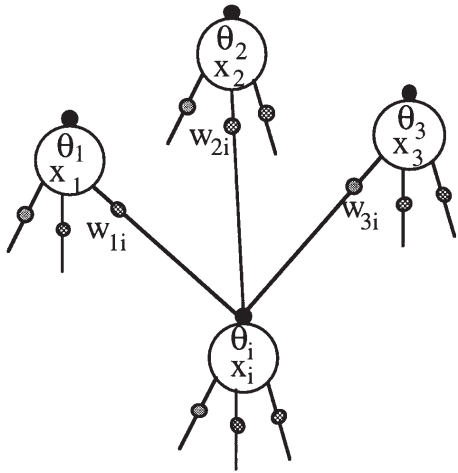,width=8truecm}
\caption []{A general connectionist network with state variables($x_{i}$), connection
strengths($W_{ij}$) and node parameters($\theta_{i}$)}
\end{center}
\end{figure}
The node parameter represents the possibility
of changing, through learning, the nature of the linear or non-linear
function $f_{i}(\sum_{j}W_{ij}x_{j})$ at each node. In the simplest case $%
\theta _{i}$ will be simply an intensity parameter. Therefore the degree to
which the activity at node $i$ influences the activity at other nodes
depends not only on the connection strengths ($W_{ij}$) but also on an
adaptive node parameter $\theta _{i}$. In some cases, as in the B-cell
immune network, the node parameter is the only means to control the relative
influence of a node on others, the connection strengths being fixed chemical
reaction rates.

\subsubsection{Hebbian - type learning with a node parameter}

We will denote by $x_{i}$ the output of node $i$. Hebbian learning \cite
{Hebb} is a type of unsupervised learning where a connection strength $%
W_{ij} $ is reinforced whenever the product $x_{i}x_{j}$ is large. As shown
by several authors, Hebbian learning extracts the eigenvectors of the
correlation matrix Q of the input data. 
\begin{equation}
Q_{ij}=\left\langle x_{i}x_{j}\right\rangle  \label{2.2}
\end{equation}
where $\left\langle ...\right\rangle $ means the sample average. If the
learning law is local, the lines of the connection matrix $W_{ij}$ all tend
to the eigenvector associated to the largest eigenvalue of the correlation
matrix. To obtain the other eigenvector directions one needs non-local laws 
\cite{Sanger} \cite{Oja1} \cite{Oja2}. Sanger's approach has the advantage
of organizing the connection matrix in such a way that the rows are the
eigenvectors associated to the eigenvalues in decreasing order. It suffers
however from slow convergence rates for the lowest eigenvalues. The methods
that have been proposed may, with small modifications, be used both for
linear and non-linear networks. However, because the maximum information
about a signal $\{x_{i}\}$, that may be coded directly in the connection
matrix $W_{ij}$ , is the principal components decomposition and this may
already be obtained with linear units, we will discuss only this case.

The learning rules proposed below\cite{Dente3} are a generalization of
Sanger's scheme including a node parameter $\theta _{i}$. We consider a
one-layer feedforward network with as many inputs as outputs (Fig.2) 
\begin{figure}[htb]
\begin{center}
\psfig{figure=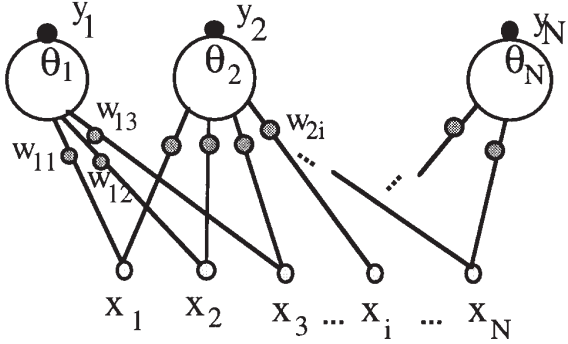,width=9truecm}
\caption []{One-layer neural feedforward network for Hebbian learning with node 
parameter}
\end{center}
\end{figure}
and the
updating rules proposed for $W_{ij}$ and $\theta _{i}$ are: 
\begin{equation}
\begin{array}{rll}
W_{ij}(t+1) & = & W_{ij}(t)+\gamma _{w}y_{i}(t)\left\{
x_{j}(t)-\sum_{k=1}^{i}\theta _{k}^{-1}y_{k}(t)W_{kj}(t)\right\} \\ 
\theta _{i}(t+1) & = & \theta _{i}(t)+\gamma _{\theta }y_{i}(t)\left\{
1-y_{i}(t)\right\}
\end{array}
\label{2.3}
\end{equation}
where $y_{i}$ is the output of node i 
\begin{equation}
y_{i}=\theta _{i}\sum_{j}W_{ij}x_{j}  \label{2.4}
\end{equation}
and $\gamma _{w}$ and $\gamma _{\theta }$ are positive constants that
control the learning rate. As will be shown below the learning dynamics of
Eqs(\ref{2.3}) has the capability to accelerate the convergence rate for the
small eigenvalues of $Q$. To avoid undesirable fixed points of the dynamics,
where this acceleration effect is not obtained, the learning rules (\ref{2.3}%
) are supplemented by the following prescription: ``{\it The starting }$%
\theta _{i}${\it 's are all positive and are not allowed to decrease below a
value (}$\theta _{i}${\it )}$_{\min }${\it . If, at a certain point of the
learning process, }$\theta _{i}${\it \ hits the lower bound then one makes
the replacement }$W_{ij}\rightarrow -W_{ij}${\it \ in the line i of the
connection matrix}`` (see remark 3).

Define the N-dimensional vectors 
\begin{equation}
\begin{array}{lll}
\left( {\bf x}\right) _{i} & = & x_{i} \\ 
\left( {\bf W}_{i}\right) _{j} & = & W_{ij}
\end{array}
\label{2.5}
\end{equation}

Then the following result was proven \cite{Dente3}:

{\it If the time scale of the system (\ref{2.3}) is much slower than the
averaging time of the input signal }$\{${\it x}$_{i}\}${\it \ then:}

{\it a) the system (\ref{2.3}) has a stable fixed point such that} 
\begin{equation}
\begin{array}{lll}
Q{\bf W}_{i} & = & \lambda _{i}{\bf W}_{i} \\ 
\theta _{i} & = & \frac{1}{\lambda _{i}}\left( {\bf W\cdot <x>}\right)
\end{array}
\label{2.6}
\end{equation}

{\it and }$\left| {\bf W}_{i}\right| =1${\it . }$\lambda _{i}${\it \ }$>0%
{\it \ }${\it (i=1...,N) are the eigenvalues of the correlation matrix.}

{\it b) Convergence to the fixed point is sequential in the sense that }$%
{\bf W}_{i}${\it \ is only attracted to the eigenvector described in (\ref
{2.6}) if all vectors }${\bf W}_{i}${\it \ for }$i<j${\it \ are already
close to their corresponding eigenvector values.}

Remarks:

1 - The matrix ${\bf W}$ of the connection strengths extracts the principal
components of the correlation matrix $Q$. The node parameters $\theta _{i}$
at their fixed point (\ref{2.6}) extract additional information on the mean
value of the data vector ${\bf x}$ and the eigenvalues of $Q$. To deal with
data with zero mean it is convenient to change the $\theta _{i}-$updating
law to 
\begin{equation}
\theta _{i}(t+1)=\theta _{i}(t)+\gamma _{\theta }\left\{ \theta
_{i}(t)\sum_{k}W_{ij}(x_{k}+r_{k})-\left( \theta
_{i}(t)\sum_{k}W_{ik}x_{k}\right) ^{2}\right\}  \label{2.13}
\end{equation}

where ${\bf r}$ is a fixed vector.

The stable fixed point is now 
\begin{equation}
\theta _{i}=\frac{1}{\lambda _{i}}\left( {\bf W}_{i}{\bf \cdot }\left( <{\bf %
x>+r}\right) \right)  \label{2.14}
\end{equation}

If one wishes to separate the eigenvalues from the information on the
average data $<{\bf x}>$ one may add another parameter $\mu _{i}$ to each
node with a learning law 
\begin{equation}
\mu _{i}(t+1)=\mu _{i}(t)+\gamma _{\mu }\left\{ 1-\mu _{i}^{\alpha
}(t)\left( \sum_{k}W_{ik}x_{k}\right) ^{2}\right\}  \label{2.15}
\end{equation}

which, with the same assumptions about time scales as before, converges to
the stable fixed point 
\begin{equation}
\mu _{i}=\left( \frac{1}{\lambda _{i}}\right) ^{\frac{1}{\alpha }}
\label{2.16}
\end{equation}

The convergence rate near the fixed point is $\gamma _{\mu }\alpha \lambda
_{i}^{\frac{1}{\alpha }}$. Therefore choosing a large $\alpha $ accelerates
convergence for the small eigenvalues.

2 - In addition to its role in extracting additional information on the
input signal, the node parameters $\theta _{i}$ also plays a role in
accelerating the convergence to the stable fixed point. For fixed $\gamma
_{w}\theta _{i}$, the rate of convergence to the fixed point is very slow
for the components associated to the smallest eigenvalues. This is the
reason for the convergence problems in Sanger's method and one also finds
that sometimes the results for the small components are quite misleading. On
the other hand to increase $\gamma _{w}$ does not help because then the time
scale of the $W_{ij}$ learning law becomes of the order of the time scale of
the data and one obtains large fluctuations in the principal components.
With a node parameter and the learning law (\ref{2.3}) the situation is more
favorable because for small eigenvalues the effective control parameter ($%
\gamma _{w}\theta _{i}$) is dynamically amplified. This accelerates the
convergence of the minor components without inducing fluctuations on the
principal components.

3- Here one examines the effect of the prescription to avoid the fixed
points where some $\theta _{i}=0$. Consider the average evolution of $\theta
_{i}$, assuming all the other variables fixed 
\[
\theta _{i}(t+1)=\theta _{i}(t)\left( 1+\gamma _{\theta }{\bf W}_{i}{\bf %
\cdot <x>}\right) -\gamma _{\theta }\theta _{i}^{2}(t)\left( {\bf W}_{i}{\bf %
\cdot QW}_{i}\right) 
\]
If ${\bf W}_{i}{\bf \cdot <x>>}0$ the stable fixed point is at ${\bf W}_{i}%
{\bf \cdot <x>/}\left( {\bf W}_{i}{\bf \cdot QW}_{i}\right) $ and if ${\bf W}%
_{i}{\bf \cdot <x><}0$ it is at zero (see Fig.3). 
\begin{figure}[htb]
\begin{center}
\psfig{figure=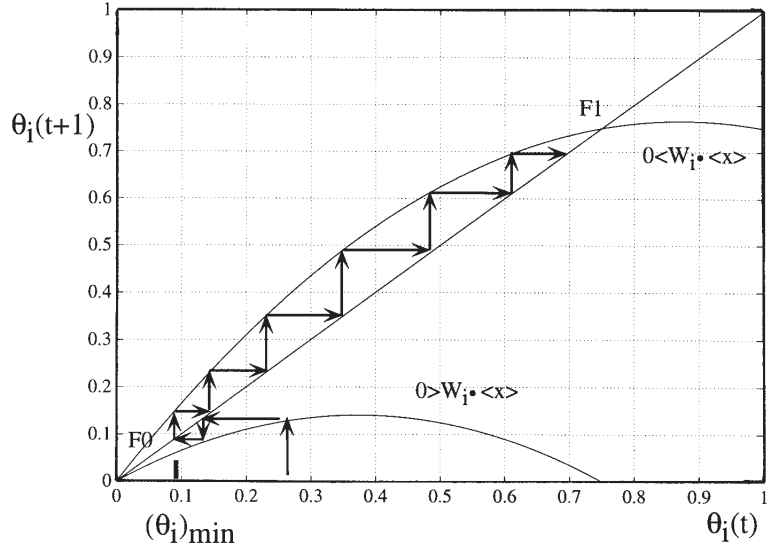,width=11truecm}
\caption []{The effect of changing the sign of $W_{i}$ in the approach to the stable 
fixed points}
\end{center}
\end{figure}
If ${\bf W}_{i}{\bf \cdot
<x>}$ is $<0$, $\theta _{i}$ moves towards the fixed point F0 at zero but,
when it reaches $\left( \theta _{i}\right) _{\min }$, the change in the sign
of the corresponding row ${\bf W}_{i}$ in connection matrix changes the
dynamics and it is F1 that is now the attracting fixed point.

To illustrate the effect of node parameters in principal component analysis
(PCA), consider the following two - dimensional $x$ signal, where $t_{i}$
are Gaussian distributed variables with zero mean and unit variance: 
\[
x_{1}=t_{1};x_{2}=t_{1}+0.08t_{2} 
\]
The principal components of the signal and the eigenvalues are given in the
following table. 
\[
\begin{tabular}{ccc}
$\lambda _{i}$ & $W_{i1}$ & $W_{i2}$ \\ 
2.0032 & 0.7060 & 0.7082 \\ 
0.0032 & 0.7082 & $-$0.7060
\end{tabular}
\]

A one - layer network with node parameters as in Fig.2 is used to perform
PCA. 
\begin{figure}[htb]
\begin{center}
\psfig{figure=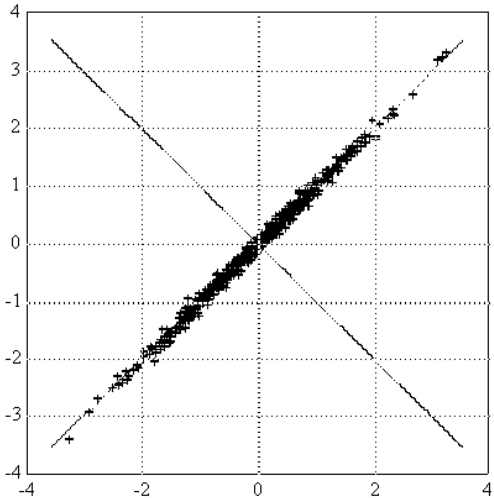,width=10truecm}
\caption []{Signal distribution and its principal directions}
\end{center}
\end{figure}
Fig.4 shows the data and the principal directions that are obtained
using the learning laws (\ref{2.3}) and (\ref{2.13}). The parameter values
used are $\gamma _{w}$ =0.015, $\gamma _{\theta }$ =0.005 and {\bf r}%
=(0.002,0.002). 
\begin{figure}[htb]
\begin{center}
\psfig{figure=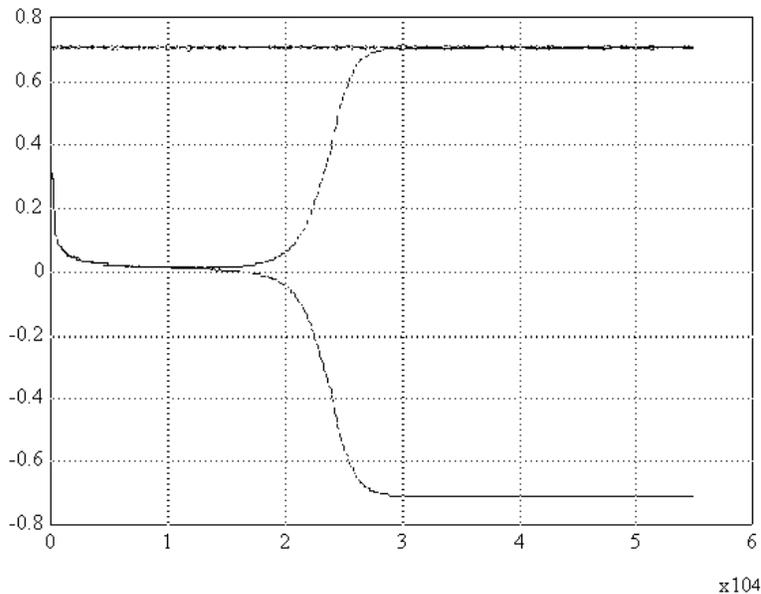,width=11truecm}
\caption []{Evolution of $W_{1}$ and $W_{2}$}
\end{center}
\end{figure}
Fig.5 shows the convergence of the ${\bf W}$'s to their
final values in the learning process. Fig.6 shows the variation of the node
parameters.
\begin{figure}[htb]
\begin{center}
\psfig{figure=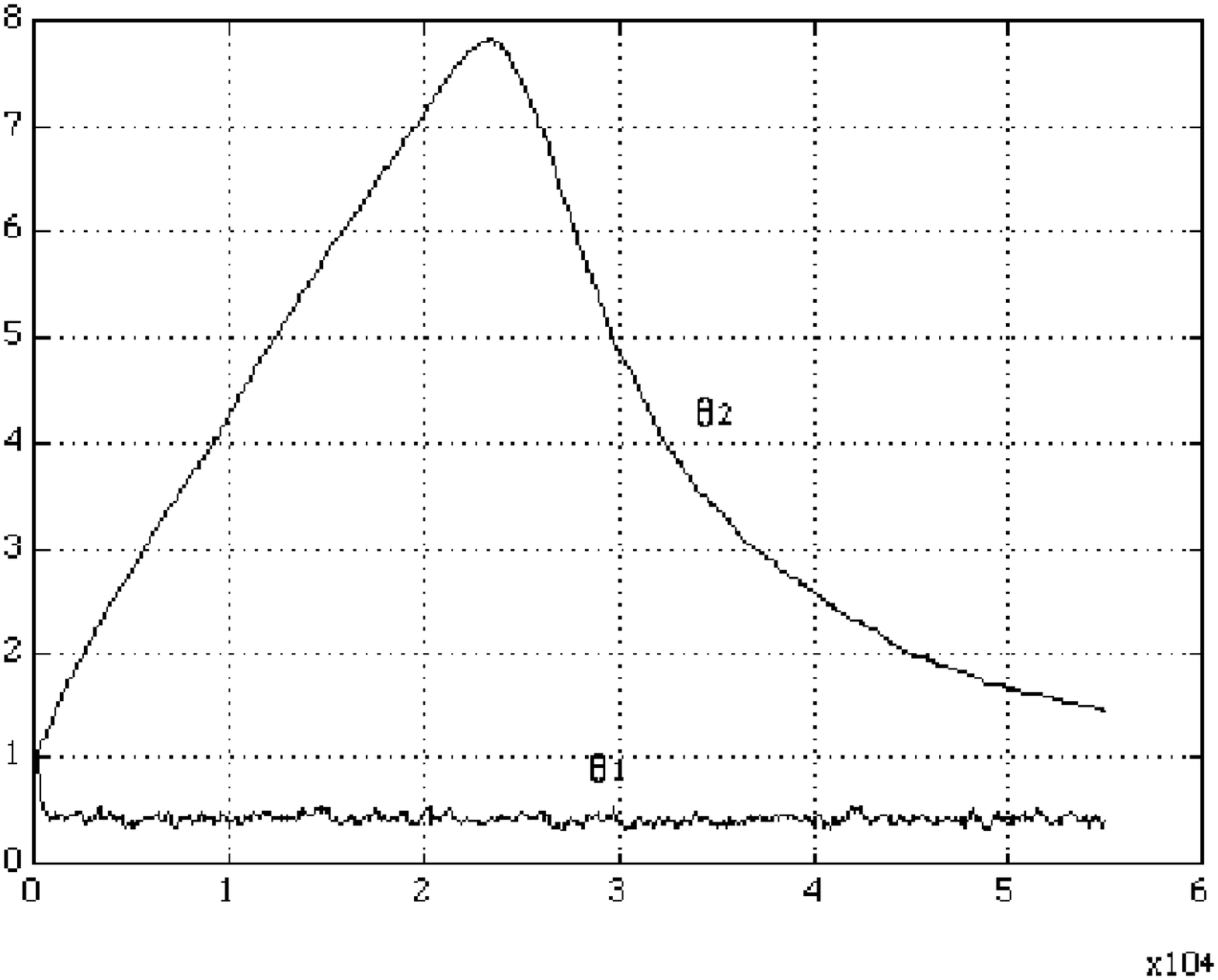,width=11truecm}
\caption []{Evolution of $\theta_{1}$ and $\theta_{2}$}
\end{center}
\end{figure}

Sanger's original algorithm is recovered by fixing the node parameters to
unit values. In this case the principal components are also extracted, but
the convergence of the process is much slower as shown in Fig.7.
\begin{figure}[htb]
\begin{center}
\psfig{figure=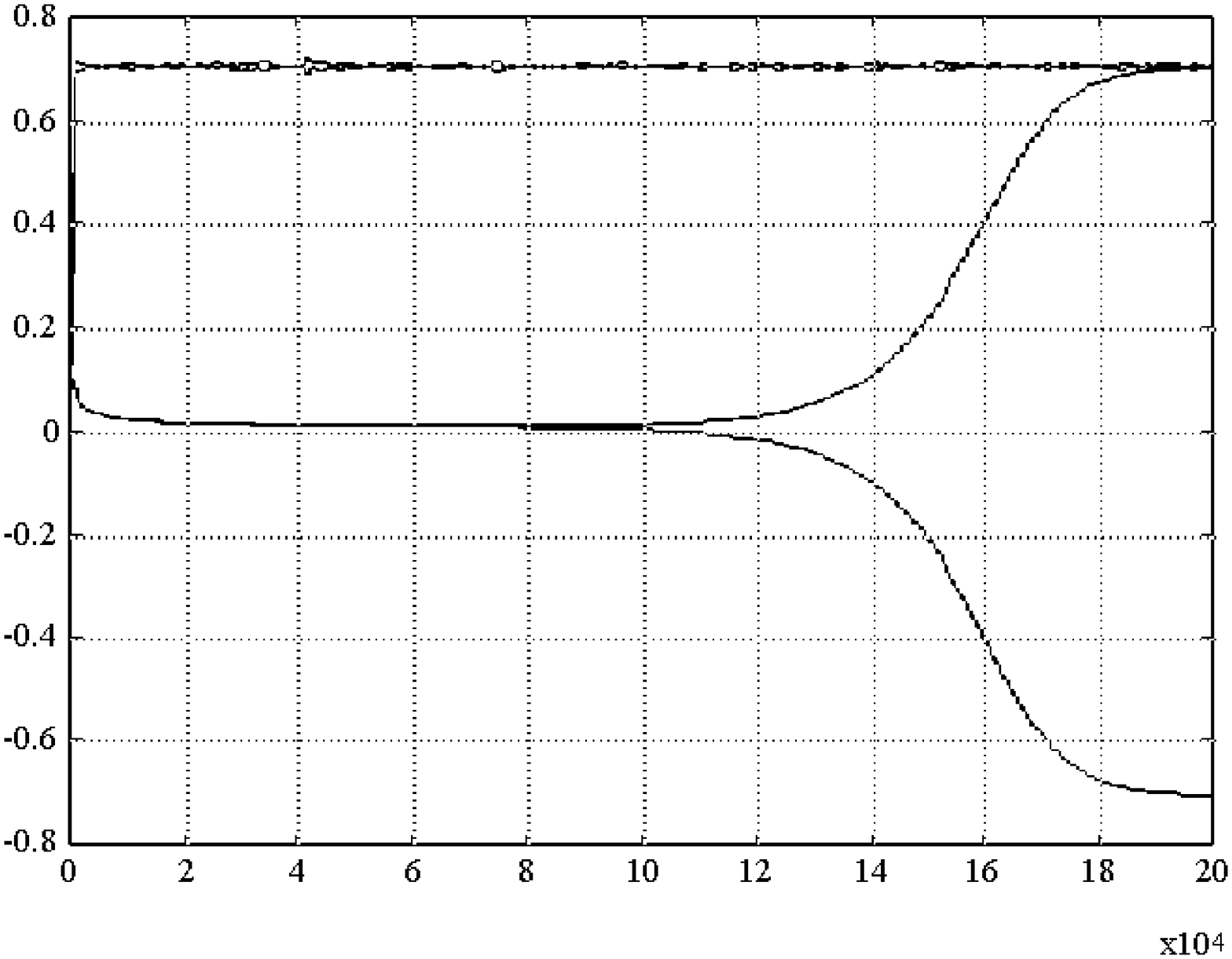,width=11truecm}
\caption []{Evolution of $W_{1}$ and $W_{2}$ using Sanger's algorithm, without
node parameters}
\end{center}
\end{figure}
 With node
parameters, improved convergence of the small eigenvalues is to be expected
under generic conditions because the rates of convergence are controlled by $%
\gamma _{w}\theta _{i}$ and $\theta _{i}$ is proportional to $\frac{1}{%
\lambda _{i}}$. In the example the effect of $\theta _{2}$ is further
amplified by the fact that, before ${\bf W}_{2}$ starts to converge, ${\bf W}%
_{2}{\bf \cdot r}$ is large. Hence, in this case at least, the node
parameter acts like a variable learning rate for the minor component. Notice
that the method does not induce oscillations in the major components as
would happen with large $\gamma _{w}$ values. Besides speeding up the
learning process the node parameters also contain information about first
moments and the eigenvalues.

Node parameters also have a beneficial effect on competitive learning
algorithms. For details refer to \cite{Dente3}.

\subsection{Non-Gaussian data and the neural computation of the
characteristic function}

The aim of principal component analysis (PCA) is to extract the eigenvectors
of the correlation matrix, from the data. There are standard neural network
algorithms for this purpose\cite{Sanger} \cite{Oja1} \cite{Oja2} \cite
{Dente3}. However if the process is non-Gaussian, PCA algorithms or their
higher-order generalizations provide only incomplete or misleading
information on the statistical properties of the data.

Let $x_{i}$ denote the output of node $i$ in a neural network. Hebbian
learning\cite{Hebb} is a type of unsupervised learning where the neural
network connection strengths $W_{ij}$ are reinforced whenever the products $%
x_{j}x_{i}$ are large. The simplest form is 
\begin{equation}
\Delta W_{ij}=\eta x_{i}x_{j}  \label{3.11}
\end{equation}
Hebbian learning extracts the eigenvectors of the correlation matrix $Q$%
\begin{equation}
Q_{ij}=<x_{i}x_{j}>  \label{3.12}
\end{equation}
but, if the learning law is local as in Eq.(\ref{3.11}), all the lines of
the connection matrix $W_{ij}$ converge to the eigenvector with the largest
eigenvalue of the correlation matrix. To obtain other eigenvector directions
requires non-local laws. These principal component analysis (PCA) algorithms
find the characteristic directions of the correlation matrix $%
(Q)_{ij}=<x_{i}x_{j}>$. If the data has zero mean ($<x_{i}>=0$) they are the
orthogonal directions along which the data has maximum variance. If the data
is Gaussian in each channel, it is distributed as a hyperellipsoid and the
correlation matrix $Q$ already contains all the information about
statistical properties. This is because higher order moments of the data may
be obtained from the second order moments. However, if the data is
non-Gaussian, the PCA analysis is not complete and higher order correlations
are needed to characterize the statistical properties. This led some authors%
\cite{Softky} \cite{Taylor} to propose networks with higher order neurons to
obtain the higher order statistical correlations of the data. An higher
order neuron is one that is capable of accepting, in each of its input
lines, data from two or more channels at once. There is then a set of
adjustable strengths $W_{ij_{1}},W_{ij_{1}j_{2}},...,W_{ij_{1}...j_{n}}$, $n$
being the order of the neuron. Networks with higher order neurons have
interesting applications, for example in fitting data to a high-dimensional
hypersurface. However there is a basic weakness in the characterization of
the statistical properties of non-Gaussian data by higher order moments.
Existence of the moments of a distribution function depends on the behavior
of this function at infinity and it frequently happens that a distribution
has moments up to a certain order, but no higher ones. A well-behaved
probability distribution might even have no moments of order higher than one
(the mean). In addition a sequence of moments does not necessarily determine
a probability distribution function uniquely\cite{Lukacs}. Two different
distributions may have the same set of moments. Therefore, for non-Gaussian
data, the PCA algorithms or higher order generalizations may lead to
misleading results.

As an example consider the two-dimensional signal shown in Fig.8.
\begin{figure}[htb]
\begin{center}
\psfig{figure=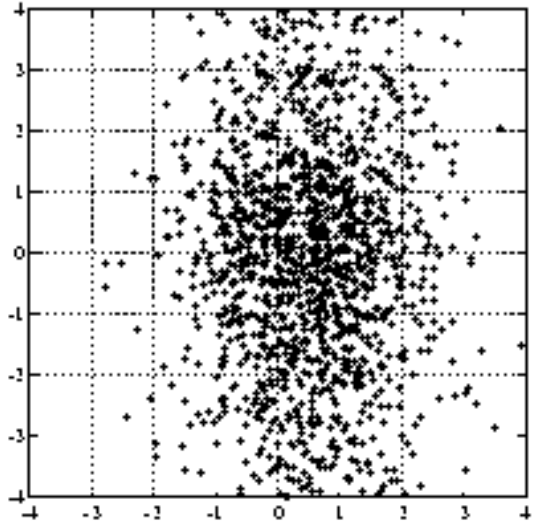,width=9truecm}
\caption []{A two-dimensional test signal}
\end{center}
\end{figure}
Fig.9 shows the evolution of the connection strengths $W_{11}$ and $W_{12}$ when
this signal is passed through a typical PCA algorithm.
\begin{figure}[htb]
\begin{center}
\psfig{figure=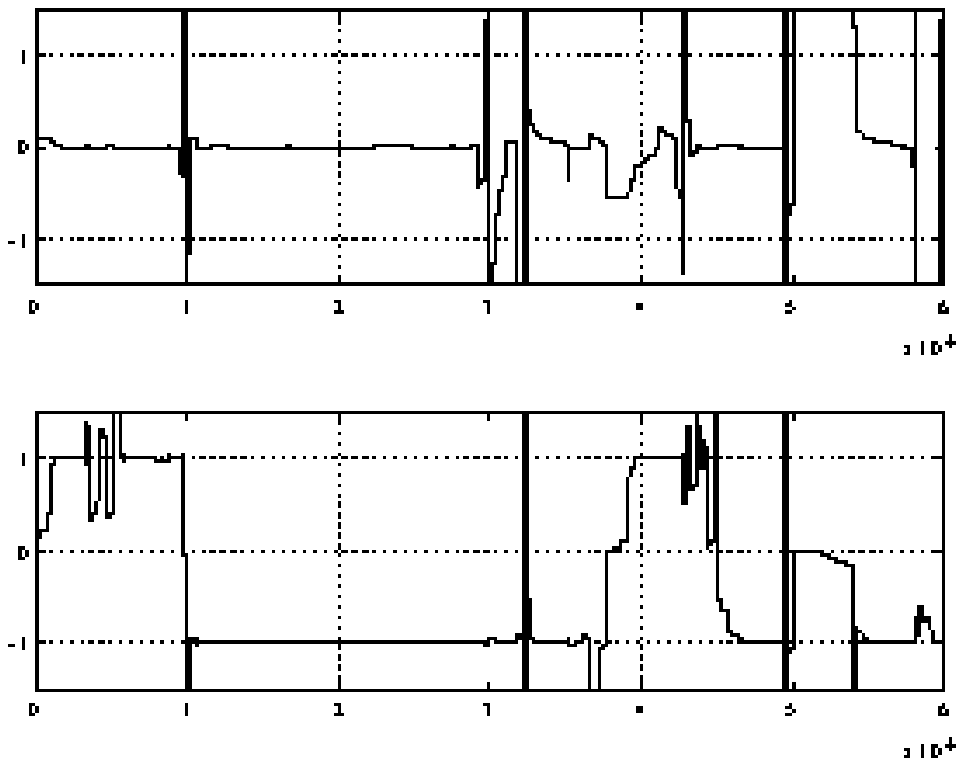,width=11truecm}
\caption []{Evolution of the connection strengths $W_{11}$ and $W_{12}$ in a PCA 
network for the data in Fig.8}
\end{center}
\end{figure}
Large oscillations
appear and finally the algorithm overflows. Smaller learning rates do not
introduce qualitative modifications in this evolution. The values may at
times appear to stabilize, but large spikes do occur. The reason is that the
seemingly harmless data in Fig.8 is generated by a linear combination of a
Gaussian with the following distribution 
\begin{equation}
p(x)=k(2+x^{2})^{-\frac{3}{2}}  \label{3.13}
\end{equation}
which has first moment, but no moments of higher order.

To be concerned with non-Gaussian processes is not a pure academic exercise,
because in many applications adequate tools are needed to analyze such
processes. For example, processes without higher order moments, in
particular those associated with L\'{e}vy statistics, are prominent in
complex processes such as relaxation in glassy materials, chaotic phase
diffusion in Josephson junctions and turbulent diffusion\cite{Shlesinger} 
\cite{Zumofen1} \cite{Zumofen2}. Moments of an arbitrary probability
distribution may not exist. However, because every bounded and measurable
function is integrable with respect to any distribution, the existence of
the characteristic function f($\alpha $) is always assured\cite{Lukacs}. 
\begin{equation}
f(\alpha )=\int e^{i\alpha .x}dF(x)=<e^{i\alpha .x}>  \label{3.14}
\end{equation}
$\alpha $ and $x$ are $N$-dimensional vectors, $x$ is the data vector and $%
F(x)$ its distribution function. The characteristic function is a compact
and complete characterization of the probability distribution of the signal.
If, in addition, one wishes to describe the time correlations of the
stochastic process $x(t)$, the corresponding quantity is the characteristic
functional\cite{Hida} 
\begin{equation}
F(\xi )=\int e^{i(x,\xi )}d\mu (x)  \label{3.15}
\end{equation}
where $\xi (t)$ is a smooth function and the scalar product is 
\begin{equation}
(x,\xi )=\int dtx(t)\xi (t)  \label{3.16}
\end{equation}
$\mu (x)$ being the probability measure over the sample paths of the process.

In the following I will describe an algorithm to compute the characteristic
function from the data, by a learning process\cite{Dente4}. The main idea is
that, in the end of the learning process, one has a neural network which is
a representation of the characteristic function. This network is then
available to provide all the required information on the probability
distribution of the data being analyzed.

Suppose we want to learn the characteristic function $f(\alpha )$ of a
one-dimensional signal $x(t)$ in a domain $\alpha \in [\alpha _{0},\alpha
_{N}]$ . The $\alpha $-domain is divided in $N$ intervals by a sequence of
values $\alpha _{0},\alpha _{1},\alpha _{2},...,\alpha _{N}$ and a network
is constructed with $N+1$ intermediate layer nodes and an output node (Fig.
10).
\begin{figure}[htb]
\begin{center}
\psfig{figure=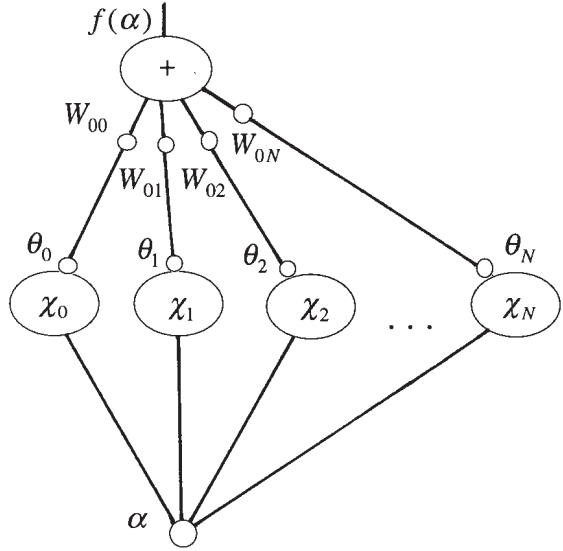,width=9truecm}
\caption []{Network to learn the characteristic function of a scalar process}
\end{center}
\end{figure}

The learning parameters in the network are the connection strengths $W_{0i}$
and the node parameters $\theta _{i}$. The existence of the node parameter
means that the output of a node in the intermediate layer is $\theta
_{i}\chi _{i}(\alpha )$, $\chi _{i}$ being a non-linear function. The use of
both connection strengths and node parameters in neural networks makes them
equivalent to a wide range of other connectionist systems\cite{Farmer} and
improves their performance in standard applications\cite{Dente3}. The
learning laws for the network in Fig. 10 are: 
\begin{equation}
\begin{array}{c}
\theta _{i}(t+1)=\theta _{i}(t)+\gamma (\cos \alpha _{i}x(t)-\theta _{i}(t))
\\ 
W_{0i}(t+1)=W_{0i}(t)+\eta \sum_{j}\left( \theta
_{j}(t)-\sum_{k}W_{0k}(t)\chi _{k}(\alpha _{j})\theta _{k}(t)\right) \theta
_{i}(t)\chi _{i}(\alpha _{j})
\end{array}
\label{3.17}
\end{equation}
$\gamma ,\eta >0$. The intermediate layer nodes are equipped with a radial
basis function 
\begin{equation}
\chi _{i}(\alpha )=\frac{e^{-(\alpha -\alpha _{i})^{2}/2\sigma _{i}^{2}}}{%
\sum_{k=0}^{N}e^{-(\alpha -\alpha _{k})^{2}/2\sigma _{k}^{2}}}  \label{3.18}
\end{equation}
where in general one uses $\sigma _{i}=\sigma $ for all $i$. The output is a
simple additive node. The learning constant $\gamma $ should be sufficiently
small to insure that the learning time is much smaller than the
characteristic times of the data $x(t)$. If this condition is satisfied each
node parameter $\theta _{i}$ tends to $<\cos \alpha _{i}x>$, the real part
of the characteristic function $f(\alpha )$ for $\alpha =\alpha _{i}$. The $%
W_{oi}$ learning law was chosen to minimize the error function 
\begin{equation}
f(W)=\frac{1}{2}\sum_{j}\left( \theta _{j}-\sum_{k}W_{0k}(t)\chi _{k}(\alpha
_{j})\theta _{k}\right) ^{2}  \label{3.19}
\end{equation}
One sees that the learning scheme is an hybrid one, in the sense that the
node parameter $\theta _{i}$ learns, in an unsupervised way, (the real part
of) the characteristic function $f(\alpha _{i})$ and then, by a supervised
learning scheme, the $W_{0i}$'s are adjusted to reproduce the $\theta _{i}$
value in the output whenever the input is $\alpha _{i}$. Through the
learning law (\ref{3.17}) each node parameter $\theta _{i}$ converges to $%
<\cos \alpha _{i}x>$ and the interpolating nature of the radial basis
functions guarantees that, after training, the network will approximate the
real part of the characteristic function for any $\alpha $ in the domain [$%
\alpha _{0},\alpha _{N}$]. A similar network is constructed for the
imaginary part of the characteristic function, where now 
\begin{equation}
\theta _{i}^{^{\prime }}(t+1)=\theta _{i}^{^{\prime }}(t)+\gamma (\sin
\alpha _{i}x(t)-\theta _{i}^{^{\prime }}(t))  \label{3.20}
\end{equation}
For higher dimensional data the scheme is similar. The number of required
nodes is $N^{d}$ for a $d$-dimensional data vector $\overrightarrow{x}(t)$.
For example for the 2-dimensional data of Fig.8 a set of $N^{2}$ nodes was
used (Fig.11).
\begin{figure}[htb]
\begin{center}
\psfig{figure=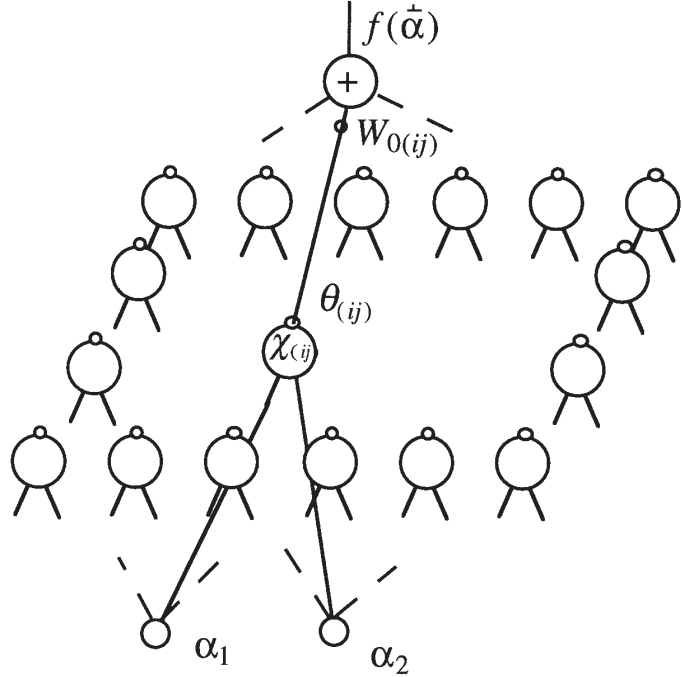,width=11truecm}
\caption []{Network to learn the characteristic function of a 2-dimensional
signal $\overrightarrow{x}(t)$}
\end{center}
\end{figure}
 Each node in the square lattice has two inputs for the two
components $\alpha _{1}$ and $\alpha _{2}$ of the vector argument of $f(%
\overrightarrow{\alpha })$. The learning laws are, as before 
\begin{equation}
\begin{array}{c}
\theta _{(ij)}(t+1)=\theta _{(ij)}(t)+\gamma (\cos \overrightarrow{\alpha }%
_{(ij)}.\overrightarrow{x}(t)-\theta _{(ij)}(t)) \\ 
W_{0(ij)}(t+1)=W_{0(ij)}(t)+ \\ 
+\eta \sum_{(kl)}\left( \theta _{(kl)}(t)-\sum_{(mn)}W_{0(mn)}(t)\chi
_{(mn)}(\overrightarrow{\alpha }_{(kl)})\theta _{(mn)}(t)\right) \theta
_{(ij)}(t)\chi _{(ij)}(\overrightarrow{\alpha }_{(kl)})
\end{array}
\label{3.21}
\end{equation}
The pair ($ij$) denotes the position of the node in the square lattice and
the radial basis function is 
\begin{equation}
\chi _{(ij)}(\overrightarrow{\alpha })=\frac{e^{-\left| \overrightarrow{%
\alpha }-\overrightarrow{\alpha }_{(ij)}\right| ^{2}/2\sigma _{(ij)}^{2}}}{%
\sum_{(kl)}e^{-\left| \overrightarrow{\alpha }-\overrightarrow{\alpha }%
_{(kl)}\right| ^{2}/2\sigma _{(kl)}^{2}}}  \label{3.22}
\end{equation}
Two networks are used, one for the real part of the characteristic function,
and another for the imaginary part with, in Eqs.(\ref{3.22}), $\cos 
\overrightarrow{\alpha }_{(ij)}.\overrightarrow{x}(t)$ replaced by $\sin 
\overrightarrow{\alpha }_{(ij)}.\overrightarrow{x}(t)$. Figs.12 and 13 show
the values computed by the algorithm for the real and imaginary parts of the
characteristic function corresponding to the two-dimensional signal in
Fig.8. 
\begin{figure}[htb]
\begin{center}
\psfig{figure=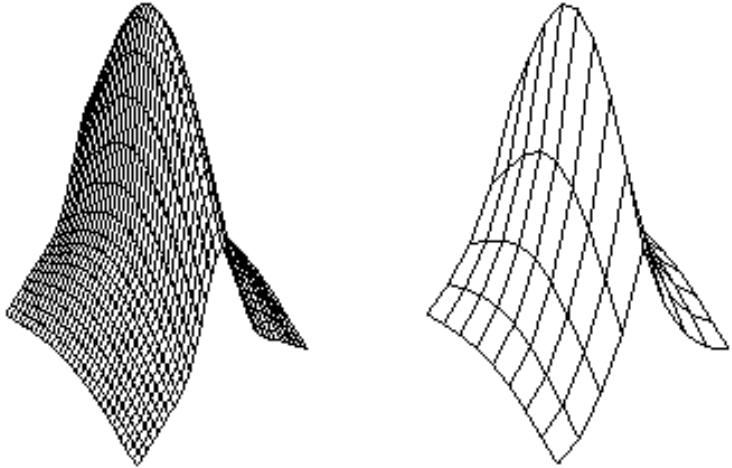,width=12truecm}
\caption []{Real part of the characteristic function for the data in Fig.8 (left) and
the mesh of $\theta_{i}$ values (right) obtained by the network}
\end{center}
\end{figure}
\begin{figure}[htb]
\begin{center}
\psfig{figure=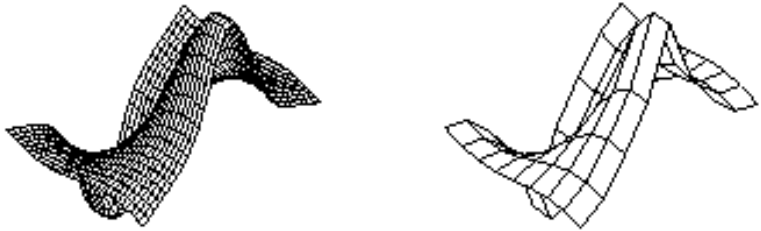,width=12truecm}
\caption []{Imaginary part of the characteristic function for the data in Fig.8 (left) and
the mesh of $\theta_{i}$ values (right) obtained by the network}
\end{center}
\end{figure}

On the left is a plot of the exact characteristic function and on the
right the values learned by the network. In this case we show only the mesh
corresponding to the $\theta _{i}$ values. One obtains a 2.0\% accuracy for
the real part and 4.5\% accuracy for the imaginary part. The convergence of
the learning process is fast and the approximation is reasonably good.
Notice in particular the slope discontinuity at the origin which reveals the
non-existence of a second moment.

For a second example the data was generated by a Weierstrass random walk
with probability distribution 
\begin{equation}
p(x)=\frac{1}{6}\sum_{j=0}^{\infty }\left( \frac{2}{3}\right) ^{j}\left(
\delta _{x,b^{j}}+\delta _{x,-b^{j}}\right)   \label{3.23}
\end{equation}
and b=1.31, which is a process of the L\'{e}vy flight type. The
characteristic function, obtained by the network, is shown in Fig. 14.
\begin{figure}[htb]
\begin{center}
\psfig{figure=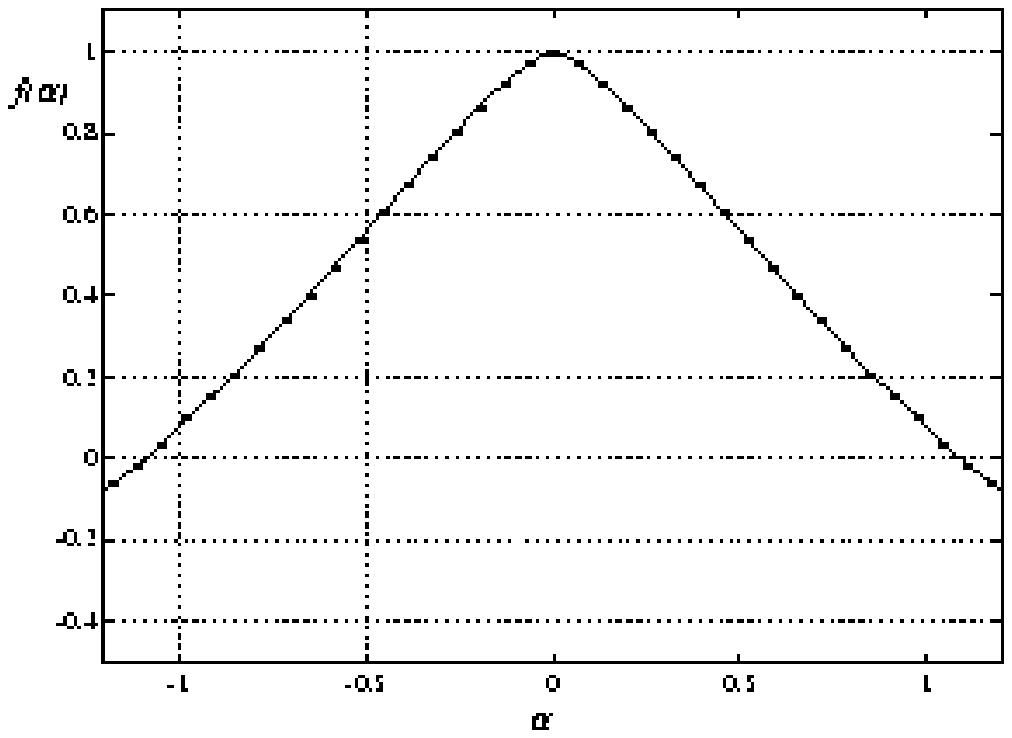,width=11truecm}
\caption []{Characteristic function for the Weierstrass random walk (b=1.31)}
\end{center}
\end{figure}

Taking the $\log (-\log )$ of the network output one obtains the scaling
exponent 1.49 near $\alpha =0$, close to the expected fractal dimension of
the random walk path (1.5).
\begin{figure}[htb]
\begin{center}
\psfig{figure=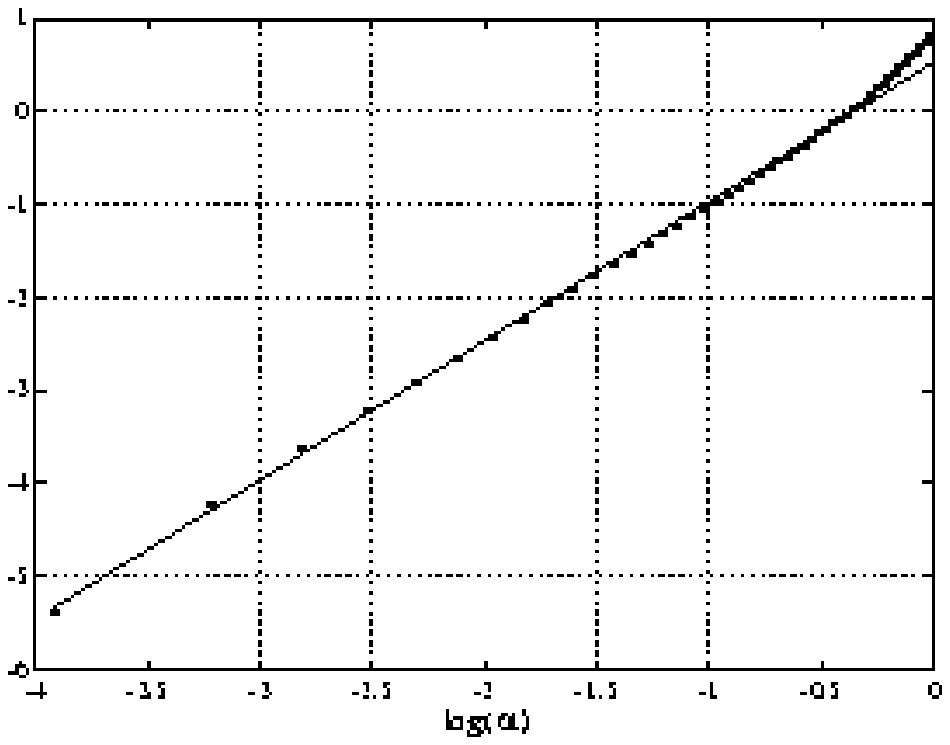,width=11truecm}
\caption []{log(-log) of the characteristic function for the Weierstrass
random walk (b=1.31)}
\end{center}
\end{figure}

These examples test the algorithm as a process identifier, in the sense
that, after the learning process, the network is a dynamical representation
of the characteristic function and may be used to perform all kinds of
analysis of the statistics of the data.

\section{Chaotic networks for information processing}

Freeman and collaborators\cite{Freeman1} \cite{Freeman2} \cite{Freeman3} 
\cite{Freeman4} \cite{Freeman5} \cite{Freeman6} have extensively studied and
modeled the neural activity in the mammalian olfactory system. Their
conclusions challenge the idea that pattern recognition in the brain is
accomplished as in an attractor neural network\cite{Amit}. Pattern
recognition in the brain is the process by which external signals arriving
at the sense organs are converted into internal meaningful states. The
studies of the excitation patterns in the olfactory bulb of the rabbit lead
to the conclusion that, at least in this biological pattern recognition
system, there is no evolution towards an equilibrium fixed point nor does it
seem to be minimizing an energy function. Other interesting conclusions of
these biological studies are:

- the main component of the neural activity in the olfactory system is
chaotic. This is also true in other parts of the brain, periodic behavior
occurring only in abnormal situations like deep anesthesia, coma, epileptic
seizures or in areas of the cortex that have been isolated from the rest of
the brain;

- the low-level chaos that exists in absence of an external stimulus is, in
the presence of a signal, replaced by bursts lasting for about 100 ms which
have different intensities in different regions of the olfactory bulb.
Olfactory pattern recognition manifests itself as a spatially coherent
pattern of intensity;

- the recognition time is very fast, in the sense that the transition
between different patterns occurs in times as short as 6 ms. Given the
neuron characteristic response times this is clearly incompatible with the
global approach to equilibrium of an attractor neural network;

- the biological measurements that have been performed do not record the
action potential of individual neurons, but the local effect of the currents
coming out of thousands of cells. Therefore the very existence of measurable
activity bursts implies a synchronization of local assemblies of many
neurons.

Freeman, Yao and Burke\cite{Freeman4} \cite{Freeman5} model the olfactory
system with a set of non-linear coupled differential equations, the coupling
being adjusted by means of an input correlation learning scheme. Each
variable in the coupled system is assumed to represent the dynamical state
of a local assembly of many neurons. Based on numerical simulations they
conjecture that olfactory pattern recognition is realized through a
multilobe strange attractor. The system would be, most of the time, in a
basal (low-activity) state, being excited to one of the higher lobes by the
external stimulus.

To compute or even prove the existence of chaotic measures in coupled
differential equation systems is a awesome task. Therefore, even if it may
be biologically accurate, the analytical model of these authors is difficult
to deal with and unsuitable for wide application in technological pattern
recognition tasks, although one such application has indeed been attempted
by the authors\cite{Freeman6}. However the idea that efficient pattern
recognition may be achieved by a chaotic system, which selects distinct
invariant measures according to the class of external stimuli, is quite
interesting and deserves further exploration. Inspired by the biological
evidence a model has been developed \cite{Dente2}, which behaves roughly as
an olfactory system (in Freeman's sense) and, at the same time, is easier to
describe and control by analytical means. To play the role of the local
chaotic assembly of neurons a Bernoulli unit is chosen. The connection
between the units is realized by linear synapses with an input correlation
learning law and the external inputs also have adjustable gains, changing as
in a biological potentiation mechanism. This last feature turns out to be
useful to enhance the novelty-filter qualities of the system.

\subsection{A network of Bernoulli units}

The network is the fully connected system shown in Fig.16.
\begin{figure}[htb]
\begin{center}
\psfig{figure=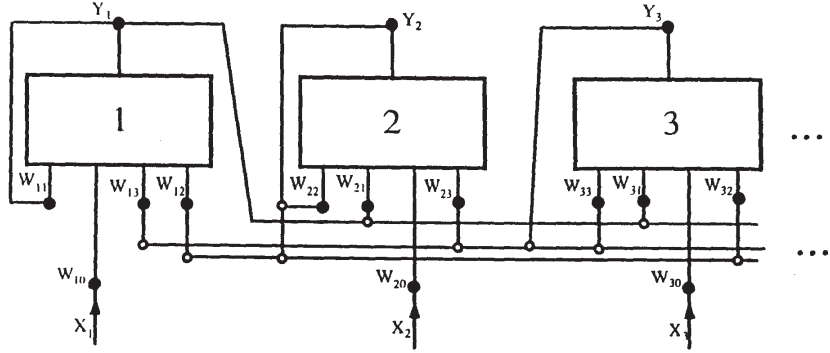,width=11truecm}
\caption []{The Bernoulli network}
\end{center}
\end{figure}
 The output of the
nodes is denoted by $y_{i}$ and the $x_{i}$'s are the external inputs. $%
W_{ij}$ with $i,j\in \{1,2,\cdots ,N\}$ are the connection strengths and $%
W_{i0}$ the input gains. Both $W_{ij}$ and $W_{i0}$ are real numbers in the
interval [0,1]. The input patterns are zero-one sequences ($x_{i}\in \{0,1\}$%
). The learning laws for the connection strengths and the input gains are
the following:

Let $S_{ij}(t)=W_{ij}(t)+\eta x_{i}(t)x_{j}(t)$ . Then for $i\neq j$%
\begin{equation}
W_{ij}(t+1)=S_{ij}(t)\textnormal{ if }\sum_{k\neq i}S_{ik}(t)\leq C  \label{3.1a}
\end{equation}
\begin{equation}
W_{ij}(t+1)=C\frac{S_{ij}(t)}{\sum_{k\neq i}S_{ik}(t)}\textnormal{ if }\sum_{k\neq
i}S_{ik}(t)>C  \label{3.1c}
\end{equation}
for $i=j$%
\begin{equation}
W_{ii}(t+1)=1-\sum_{j\neq i}W_{ij}(t+1)  \label{3.2}
\end{equation}
and for the input gains 
\begin{equation}
W_{i0}(t+1)=\alpha \frac{N_{i}(1)_{t}}{N_{i}(1)_{t}+N_{i}(0)_{t}}
\label{3.3}
\end{equation}

According to Eqs.(\ref{3.1a}-\ref{3.3}), when an input pattern has a one in
both the i and j positions, the correlation of the units i and j becomes
stronger. $C<1$ is a constant related to the node dynamics, which the sum of
the off-diagonal connections is not allowed to exceed. $\eta $ is a small
parameter that controls the learning speed. Finally the diagonal element $%
W_{ii}$ is chosen in such a way that the sum of all connections entering
each unit adds to one.

In the input gain learning law, $N_{i}(1)_{t}$ (or $N_{i}(0)_{t}$) is the
number of times that a one (or a zero) has appeared at the input i, up to
time t. Eq.(\ref{3.3}) means that if an input is excited many times, during
the learning phase, it becomes more sensitive.

The node dynamics is 
\begin{equation}
y_{i}(t+1)=f\left( \sum_{j}W_{ij}(t)y_{j}(t)+W_{i0}x_{i}(t)\right) 
\label{3.4}
\end{equation}
$f$ being the function depicted in Fig.17.
\begin{figure}[htb]
\begin{center}
\psfig{figure=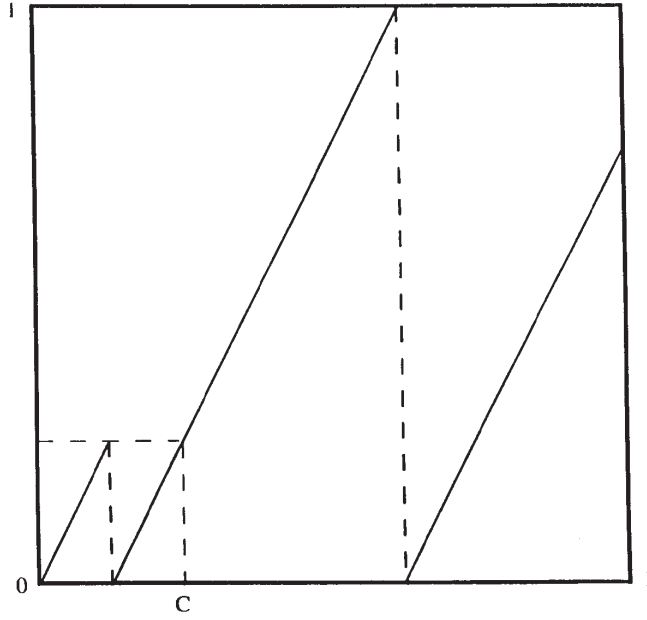,width=10truecm}
\caption []{The node dynamics function}
\end{center}
\end{figure}

The learning process starts with $W_{ii}=1$ and $W_{ij}=0$ for $i\neq j$.
Each unit has then an independent absolutely continuous invariant measure
which is the Lebesgue measure in [$0,C$] and zero outside. When the $W_{ij}$
($i\neq j$) become different from zero but the inputs $x_i$ are still zero,
all variables $y_i$ stay in the interval [$0,C$] because of the convex
linear combination of inputs imposed by the normalization of the $W_{ij}$'s.
When some inputs $x_i$ are $\neq 0$, there is a finite probability for an
irregular burst in the interval [$C,1$], of some of the node variables, with
reinjection into [$0,C$] whenever the iterate falls on the interval [$\frac
12+\frac C2,\frac 12+C$].

The bursts in the interval [$C,1$], in response to some of the input
patterns, is the recognition mechanism of the network. The basal chaotic
dynamics insuring an uniform covering of the interval [$0,C$], the timing of
the onset of the bursts depends only on the correlation probability and on
the clock time of the discrete dynamics. We understand therefore why a
chaos-based network may have a recognition time faster than an attractor
network.

{\bf Learning, invariant measures and simulations}

Both the connection strengths and the nature of the bursts, for a given set
of $W$'s and an applied input pattern, may be estimated in probability.

In Eqs.(\ref{3.1a}-\ref{3.1c}), either the node i is not correlated to any
other node and then all off-diagonal elements $W_{ij}$ are zero or, as soon
as the input patterns begin to correlate the node i with any other node, the
off-diagonal elements start to grow and only the second case (Eq.(\ref{3.1c}%
)) needs to be considered. Let the learning gain be small. Then, in first
order in $\eta $, we have 
\[
W_{ij}(t+1)=W_{ij}(t)+\eta x_{i}(t)x_{j}(t)-\frac{\eta }{C}%
W_{ij}(t)\sum_{k\neq i}x_{i}(t)x_{k}(t) 
\]
For N learning steps, in first order in $\eta $%
\[
W_{ij}(t+N)=W_{ij}(t)+\eta \sum_{n=0}^{N-1}x_{i}(t+n)x_{j}(t+n)-\frac{\eta }{%
C}W_{ij}(t)\sum_{k\neq i}\sum_{n=0}^{N-1}x_{i}(t+n)x_{k}(t+n) 
\]
Denoting by $p_{ij}(1)$ the probability for the occurrence, in the input
patterns, of a one both in the i and the j positions, the above equation has
the stationary solution 
\[
W_{ij}=C\frac{p_{ij}(1)}{\sum_{k\neq i}p_{ik}(1)} 
\]

Now we establish an equation for the burst probabilities. Consider the case
where $C$ is much smaller than one, that is, the basal chaos is of low
intensity. In this case, because of the normalization chosen for $W_{ii}$,
the dynamics inside the interval [$0,C$] is dominated by $y\rightarrow 2y$%
(mod $C$) and in the interval [$C,1$] by $y\rightarrow 2y$(mod $1$). Hence,
to a good approximation, we may assume uniform probability measures for the
motion inside each one of the intervals. Denoting the interval [$0,C$] as
the state 1 and the interval [$C,1$] as the state 2, the dynamics of each
node is a two-state Markov process with transition probabilities between the
states corresponding to the probabilities of falling in some subintervals of
the intervals [$0,C$] and [$C,1$]. Namely, the probability $p(2\rightarrow
1) $ equals the probability of falling in the reinjection interval [$\frac
12+\frac C2,\frac 12+C$] and the probability $p(1\rightarrow 2)$ that of
falling near the point $C$ at a distance smaller than the off-diagonal
excitation. 
\begin{equation}  \label{3.2a}
p_i(2\rightarrow 1)_t=\frac C{2(1-C)}
\end{equation}
\begin{equation}  \label{3.2b}
p_i(2\rightarrow 2)_t=1-p_i(2\rightarrow 1)_t
\end{equation}
\begin{equation}  \label{3.2c}
p_i(1\rightarrow 2)_t=\left\{ \frac 1{W_{ii}C}\left(
W_{i0}x_i(t)-C(1-W_{ii})+\sum_{j\neq i}W_{ij}y_j(t)\right) \right\} ^{\#}
\end{equation}
\begin{equation}  \label{3.2d}
p_i(1\rightarrow 1)_t=1-p_i(1\rightarrow 2)_t
\end{equation}
where we have used the notation 
\[
f^{\#}=(f\vee 0)\wedge 1 
\]
for functions truncated to the range [$0,1$]. That is, $f^{\#}=0$ if $f<0$, $%
f^{\#}=1$ if $f>1$ and $f^{\#}=f$ if $1\geq f\geq 0$.

The sum in the right-hand side of Eq.(\ref{3.2c}) is approximated in
probability by 
\begin{equation}
\sum_{j\neq i}W_{ij}\left( \frac{1}{2}p_{j}(2)_{t}+\frac{C}{2}%
(1-p_{j}(2)_{t})\right)  \label{3.5}
\end{equation}
where $p_{j}(2)_{t}$ denotes the probability of finding the node j in the
state 2 at time t. The probability estimate (\ref{3.5}), for the outputs $%
y_{j}$ , assumes statistical independence of the units. This hypothesis
fails when there are synchronization effects, which are to be expected
mainly when a small group of units is strongly correlated.

With the probability estimate for the $y_{j}$'s and the detailed balance
principle it is now possible to write a self-consistent equation for the
probability $p_{i}(2)_{t}$ to find an arbitrary node i in the state 2 at
time t 
\begin{equation}
p_{i}(2)_{t}=\frac{\left\{ \frac{1}{W_{ii}C}\left(
W_{i0}x_{i}(t)-C(1-W_{ii})+\sum_{j\neq i}W_{ij}\left( \frac{1}{2}%
p_{j}(2)_{t}+\frac{C}{2}(1-p_{j}(2)_{t})\right) \right) \right\} ^{\#}}{%
\left\{ \frac{1}{W_{ii}C}\left( W_{i0}x_{i}(t)-C(1-W_{ii})+\sum_{j\neq
i}W_{ij}\left( \frac{1}{2}p_{j}(2)_{t}+\frac{C}{2}(1-p_{j}(2)_{t})\right)
\right) \right\} ^{\#}+\frac{C}{2(1-C)}}  \label{3.6}
\end{equation}

For each input pattern $x_{i}(t)$, one obtains an estimate for $p_{i}(2)_{t}$
solving Eq.(\ref{3.6}) by iteration. We find that the solution that is
obtained is qualitatively similar to the numerically determined invariant
measures, although it tends to overestimate the burst excitation
probabilities when they are small. This may be understood from the
synchronization effects between groups of units. When one unit is not
excited (not in state 2) the others tend also not to be excited, hence (\ref
{3.5}) overestimates the sum $\sum_{j\neq i}W_{ij}y_{j}(t)$.

We now illustrate how the network behaves as an associator and pattern
recognizer. Consider, for display simplicity, a network of four nodes that
is exposed during many iterations to the patterns 1000 and 0110 where the
first pattern appears twice as much as the second. After this learning
period we have exposed the network to all possible zero-one input patterns
for 500 time steps each and observed the network reaction. During the recall
experiment no further adjustment of the $W_{ij}$'s is made. The result is
shown in the Fig.18. 
\begin{figure}[htb]
\begin{center}
\psfig{figure=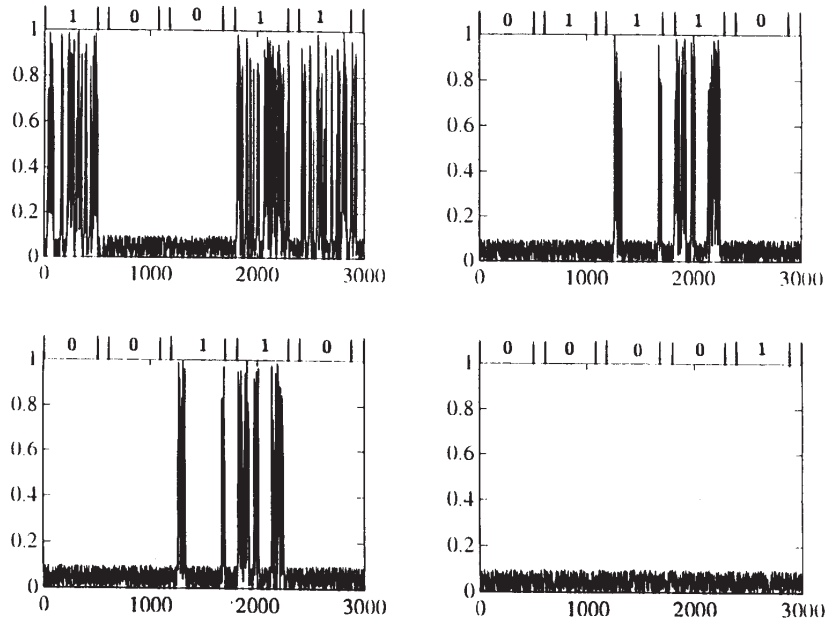,width=12truecm}
\caption []{Response of a 4-nodes network after being exposed to the patterns
1000 and 0110 ($C=0.1, \alpha=0.001$)}
\end{center}
\end{figure}

The conclusions from this and other simulations is
that, according to nature of the learning patterns, the network acts, for
the recall input patterns, as a mixture of memory, associator and novelty
filter. For example in Fig.18 we see that after having learned the sequences
1000 and 0110, the network reproduces these patterns as a memory. The
pattern 1001 is associated to the pattern 1000 and the pattern 1110
associated to a mixture of the two learned patterns. By contrast the pattern
0100 is not recognized by the network which acts then as a novelty filter.
Fig.19 shows the invariant measures of the system (expressed in probability
per bin) when the input patterns are 1100 and 0110.
\begin{figure}[htb]
\begin{center}
\psfig{figure=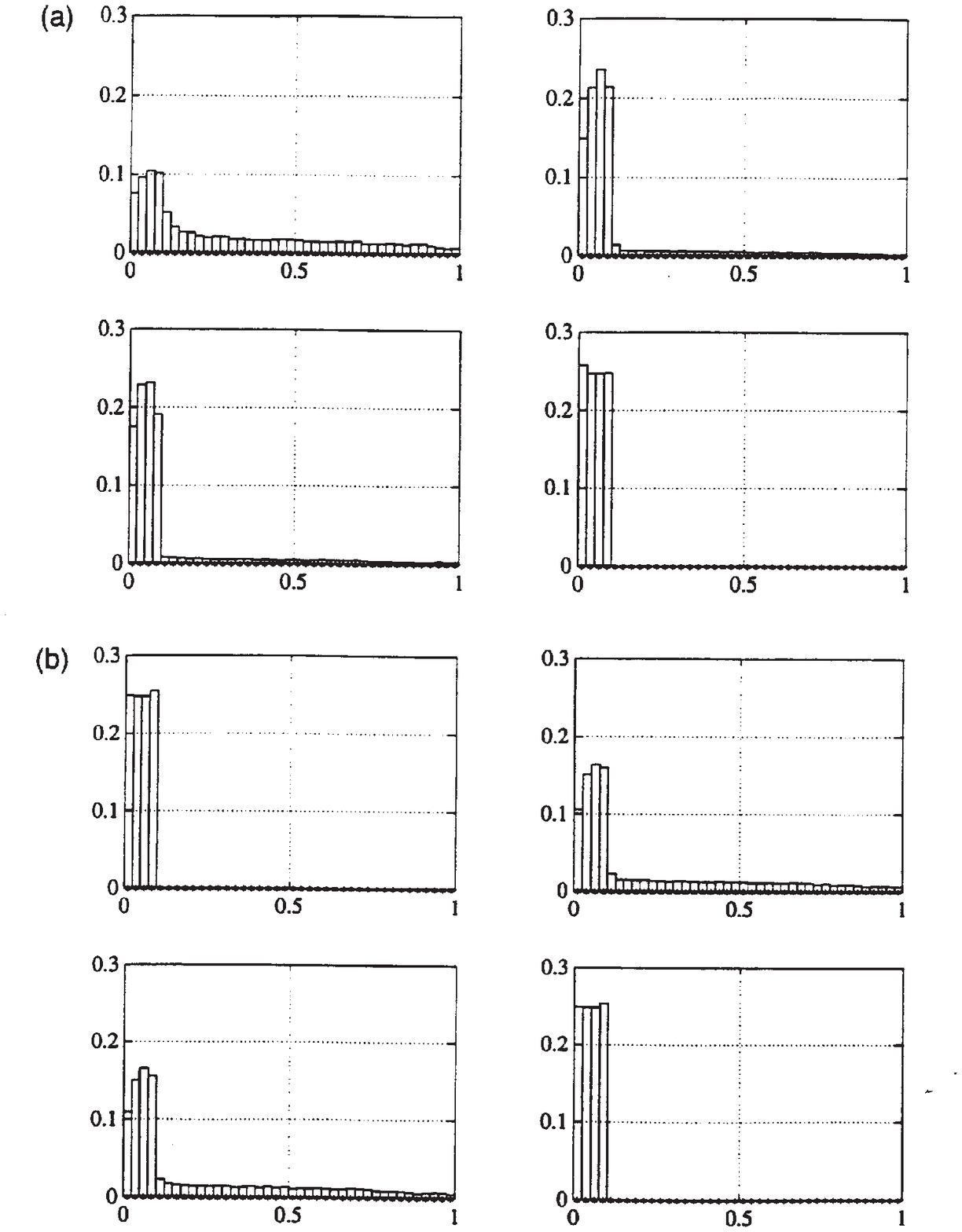,width=13truecm}
\caption []{Invariant measures for the input patterns (a)1000 and (b)0110}
\end{center}
\end{figure}

In conclusion, the network based on Bernoulli units, with node dynamics as
in Fig.17 and correlation learning described by Eqs.(\ref{3.1a}-\ref{3.3}):

1. Is a chaos-based pattern recognizer with the capability of operating on
distinct invariant measures which are selected by the input patterns;

2. The response time of the selection is controlled by the magnitude of the
invariant measures and the clock time of the basal chaotic dynamics;

3. As a pattern recognizer the network is a mixture of memory, associator
and novelty filter. This however is sensitive to the learning algorithm that
is chosen and, for the algorithm that is discussed here, it is sensitive
also to value of the parameters $C$ and $\eta $.

\subsection{Feigenbaum networks}

In this part one studies dynamical systems composed of a set of coupled
quadratic maps 
\begin{equation}
x_{i}(t+1)=1-\mu _{*}\{\sum_{j}W_{ij}x_{j}(t)\}^{2}  \label{f1}
\end{equation}
with $x\in [-1,1]$, $\sum_{j}W_{ij}=1$ $\forall i$ and $W_{ij}>0$ $\forall
i,j$ and $\mu _{*}=1.401155...$. The value chosen for $\mu _{*}$ implies
that, in the uncoupled limit ($W_{ii}=1,W_{ij}=0$ $i\neq j$), each unit
transforms as a one-dimensional quadratic map in the accumulation point of
the Feigenbaum period-doubling bifurcation cascade. This system will be
called a Feigenbaum network.

The quadratic map at the Feigenbaum accumulation point is not in the class
of chaotic systems (in the sense of having positive Lyapunov exponents)
however, it shares with them the property of having an infinite number of
unstable periodic orbits. Therefore, before the interaction sets in, each
elementary map possesses an infinite diversity of potential dynamical
behaviors. As we will show later, the interaction between the individual
units is able to selectively stabilize some of the previously unstable
periodic orbits. The selection of the periodic orbits that are stabilized
depends both on the initial conditions and on the intensity of the
interaction coefficients $W_{ij}$. As a result Feigenbaum networks appear as
systems with potential applications in the fields of control of chaos,
information processing and as models of self-organization.

Control of chaos or of the transition to chaos has been, in recent years, a
very active field (see for example Ref.\cite{Shinbrot} and references
therein). Several methods were developed to control the unstable periodic
orbits that are embedded within a chaotic attractor. Having a way to select
and stabilize at will these orbits we would have a device with infinite
storage capacity (or infinite pattern discrimination capacity). However, an
even better control might be achieved if, instead of an infinite number of
unstable periodic orbits, the system possesses an infinite number of
periodic attractors. The basins of attraction would evidently be small but
the situation is in principle more favorable because the control need not be
as sharp as before. As long as the system is kept in a neighborhood of an
attractor the uncontrolled dynamics itself stabilizes the orbit.

The creation of systems with infinitely many sinks near an homoclinic
tangency was discovered by Newhouse \cite{Newhouse} and later studied by
several other authors \cite{Robinson} \cite{Gambaudo} \cite{Tedeschini} \cite
{Wang} \cite{Nusse}. In the Newhouse phenomenon infinitely many attractors
may coexist but only for special parameter values, namely for a residual
subset of an interval. Another system, different from the Newhouse
phenomena, which also displays many coexisting periodic attractors is a
rotor map with a small amount of dissipation\cite{Feudel}.

Here one shows that for a Feigenbaum system with only two units and
symmetrical couplings one obtains a system which has an infinite number of
sinks for an open set of coupling parameters. Then one also analyzes the
behavior of a Feigenbaum network in the limit of a very large number of
units. A mean field analysis shows how the interaction between the units may
generate distinct periodic orbit patterns throughout the network.

\subsubsection{A simple system with an infinite number of sinks}

Consider two units with symmetrical positive couplings ($W_{12}=W_{21}=c>0$)

\begin{equation}
\begin{array}{c}
x_{1}(t+1)=1-\mu _{*}\left( (1-c)x_{1}(t)+cx_{2}(t)\right) ^{2} \\ 
x_{2}(t+1)=1-\mu _{*}\left( cx_{1}(t)+(1-c)x_{2}(t)\right) ^{2}
\end{array}
\label{f2.1}
\end{equation}

The mechanism leading to the emergence of periodic attractors from a system
that, without coupling, has no stable finite-period orbits is the permanence
of the unstabilized orbits in a flip bifurcation and the contraction effect
introduced by the coupling. The structure of the basins of attraction is
also understood from the same mechanism. The result is\cite{Carvalho}:

{\it For sufficiently small }$c${\it \ there is an }$N${\it \ such that the
system (\ref{f2.1}) has stable periodic orbits of all periods }$2^{n}${\it \
for }$n>N${\it .}

\subsubsection{Feigenbaum networks with many units}

{\bf Mean-field analysis}

For the calculations below it is convenient to use as variable the net input
to the units, $y_{i}=$ $\sum_{j}W_{ij}x_{j}$. Then Eq.(\ref{f1}) becomes, 
\begin{equation}
y_{i}(t+1)=1-\mu _{*}\sum_{j=1}^{N}W_{ij}y_{j}^{2}(t)  \label{f3.1}
\end{equation}

For practical purposes some restrictions have to be put on the range of
values that the connection strengths may take. For information processing
(pattern storage and pattern recognition) it is important to preserve, as
much as possible, the dynamical diversity of the system. That means, for
example, that a state with all the units synchronized is undesirable insofar
as the effective number of degrees of freedom is drastically reduced. From 
\begin{equation}
\delta y_{i}(t+1)=-2\mu _{*}y(t)(W_{ii}\delta y_{i}(t)+\sum_{j\neq
i}W_{ij}\delta y_{j}(t))  \label{f3.2}
\end{equation}
one sees that instability of the fully synchronized state implies $\left|
2\mu _{*}y(t)W_{ii}\right| >1$. Therefore, the interesting case is when the
off-diagonal connections are sufficiently small to insure that 
\begin{equation}
W_{ii}>\frac{1}{\mu _{*}}  \label{f3.3}
\end{equation}

For large $N$, provided there is no large scale synchronization effect, a
mean-field analysis might be appropriate, at least to obtain qualitative
estimates on the behavior of the network. For the unit $i$ the average value 
$<1-\mu _{*}\sum_{j\neq i}W_{ij}y_{j}^{2}(t)>$ acts like a constant and the
mean-field dynamics is 
\begin{equation}
z_{i}(t+1)=1-\mu _{i,eff}\bigskip \ z_{i}^{2}(t)  \label{f3.4}
\end{equation}
where 
\begin{equation}
z_{i}=\frac{y_{i}}{<1-\mu _{*}\sum\limits_{j\neq i}W_{ij}y_{j}^{2}>}
\label{f3.5}
\end{equation}
and 
\begin{equation}
\mu _{i,eff}=\mu _{*}W_{ii}<1-\mu _{*}\sum\limits_{j\neq i}W_{ij}y_{j}^{2}>
\label{f3.6}
\end{equation}
$\mu _{i,eff}$ is the effective parameter for the mean-field dynamics of
unit $i$. From (\ref{f3.3}) and (\ref{f3.6}) it follows $\mu _{*}W_{ii}>\mu
_{i,eff}>\mu _{*}W_{ii}(2-\mu _{*})$. The conclusion is that the effective
mean-field dynamics always corresponds to a parameter value below the
Feigenbaum accumulation point, therefore, one expects the interaction to
stabilize the dynamics of each unit in one of the $2^{n}$- periodic orbits.
On the other hand to keep the dynamics inside an interesting region we
require $\mu _{i,eff}>\mu _{2}=1.3681$, the period-2 bifurcation point. With
the estimate $<y^{2}>=\frac{1}{3}$ one obtains 
\begin{equation}
\mu _{*}W_{ii}(1-\frac{\mu _{*}}{3}(1-W_{ii}))>\mu _{2}  \label{f3.7}
\end{equation}
which, together with (\ref{f3.3}), defines the interesting range of
parameters for $W_{ii}=1-\sum\limits_{j\neq i}W_{ij}$.

{\bf Feigenbaum networks as signal processors}

Let, for example, the $W_{ij}$ connections be constructed from an input
signal $x_{i}$ by a correlation learning process 
\begin{equation}
\begin{array}{c}
W_{ij}\rightarrow W_{ij}^{^{\prime }}=(W_{ij}+\eta x_{i}x_{j})e^{-\gamma }%
\textnormal{ for }i\neq j \\ 
W_{ii}\rightarrow W_{ii}^{^{\prime }}=1-\sum_{j\neq i}W_{ij}
\end{array}
\label{f3.8}
\end{equation}
The dynamical behavior of the network, at a particular time, will reflect
the learning history, that is, the data regularities, in the sense that $%
W_{ij}$ is being structured by the patterns that occur more frequently in
the data. The decay term $e^{-\gamma }$ insures that the off-diagonal terms
remain small and that the network structure is determined by the most
frequent recent patterns. Alternatively, instead of the decay term, we might
use a normalization method and the connection structure would depend on the
weighted effect of all the data.

In the operating mode described above the network acts as a {\it signal
identifier}. For example if the signal patterns are random, there is little
correlation established and all the units operate near the Feigenbaum point.
Alternatively the learning process may be stopped at a certain time and the
network then used as a {\it pattern recognizer}. In this latter mode,
whenever the pattern \{$x_{i}$\} appears, one makes the replacement 
\begin{equation}
\begin{array}{c}
W_{ij}\rightarrow W_{ij}^{^{\prime }}=W_{ij}x_{i}x_{j}\textnormal{ for }i\neq j \\ 
W_{ii}\rightarrow W_{ii}^{^{\prime }}=1-\sum_{j\neq i}W_{ij}
\end{array}
\label{f3.9}
\end{equation}
Therefore if $W_{ij}$ was $\neq 0$ but either $x_{i}$ or $x_{j}$ is $=0$
then $W_{ij}^{^{\prime }}=0$. That is, the correlation between node $i$ and $%
j$ disappears and the effect of this connection on the lowering of the
periods vanishes.

If both $x_{i}$ and $x_{j}$ are one, then $W_{ij}^{^{\prime }}=W_{ij}$ and
the effect of this connection persists. Suppose however that for all the $%
W_{ij}$'s different from zero either $x_{i}$ or $x_{j}$ are equal to zero.
Then the correlations are totally destroyed and the network comes back to
the uncorrelated (nonperiodic behavior). This case is what is called a {\it %
novelty filter}. Conversely, by displaying periodic behavior, the network 
{\it recognizes} the patterns that are similar to those that, in the
learning stage, determined its connection structure. Recognition and {\it %
association} of similar patterns is then performed \cite{Carvalho}.

\end{document}